%% file: perturbation.tex
\documentclass[aps, amsmath,floatfix,twocolumn,prb,superscriptaddress]{revtex4-2}
\usepackage{amssymb,amsmath,amsthm,amsfonts}
\usepackage{graphicx,array}
\usepackage{dcolumn}
\usepackage{psfrag}
\usepackage{bm}
\usepackage{color}
\usepackage{hyperref}
\usepackage{tabularx}
\usepackage{array}
\usepackage{svg}
\usepackage{natbib}
\setcitestyle{numbers} 
\setcitestyle{square}
\usepackage{hyperref}
\hypersetup{
	colorlinks=true,
 	linkcolor=blue,
 	citecolor = blue,      
 	urlcolor=blue,
 }     
     
\begin{document}
\input{defs}

\title{Potential energy surface prediction of Alumina polymorphs using graph neural network}

\author{Soumya Sanyal}
\thanks{Work done while at Indian Institute of Science Bangalore}
\affiliation{University of Southern California, USA} 

\author{Arun Kumar Sagotra}
\thanks{Corresponding Author and Work done while at Shell Technology Centre Bangalore.}
\email{arunsagotra91@gmail.com and janakiraman.balachandran@gmail.com}
\affiliation{}

\author{Narendra Kumar}
\affiliation{Jawaharlal Nehru Centre for Advanced Scientific Research, Bangalore}

\author{Sharad Rathi}
\affiliation{Vellore Institute of Technology, Vellore}

\author{Mohana Krishna}
\affiliation{Vellore Institute of Technology, Vellore}

\author{Nagesh Somayajula}
\affiliation{Centura, Colarado, USA}

\author{Duraivelan Palanisamy}
\affiliation{Shell Technology Centre Bangalore, Bangalore}

\author{Ram R. Ratnakar}
\affiliation{Shell Technology Centre Houston, USA}

\author{Suchismita Sanyal}
\affiliation{Shell Technology Centre Bangalore, Bangalore}

\author{Partha Talukdar}
\thanks{Work done while at Indian Institute of Science Bangalore}
\affiliation{Google Research, Bangalore, India}

\author{Umesh Waghmare}
\affiliation{Jawaharlal Nehru Centre for Advanced Scientific Research, Bangalore}

\author{Janakiraman Balachandran}
\thanks{Corresponding Author and Work done while at Shell Technology Centre Bangalore.}
\email{arunsagotra91@gmail.com and janakiraman.balachandran@gmail.com}
\affiliation{}

\maketitle

{\color{black}\bf The  process of design and discovery of new materials can be significantly expedited and simplified if we can learn effectively from available data. Deep learning (DL) approaches have recently received a lot of interest for their ability to speed up the design of novel materials by predicting material properties with precision close to experiments and ab-initio calculations. The application of deep learning to predict materials properties measured by experiments are valuable yet challenging due to the limited amount of experimental data. Most of the existing approaches to predict properties from computational data have
also been directed towards specific material properties. In this work, we extend this approach, by proposing \methodfull{(\method{})},  an accurate and transferable deep learning framework based on graph convolutional networks. \method{} directly learns the potential energy surface (PES) from atomic configurations. This approach can enable transferable models that can predict different material properties. We apply this framework  to bulk crystals (i.e. Al$_{2}$O$_{3}$), and test it by calculating potential energy surfaces at different temperatures and across  different phases of crystal.}\\

\input{sections/introduction}
\input{sections/outline}
\input{sections/method}
\input{sections/results}
\input{sections/conclusion}
\input{sections/acknowledgement}

\bibliographystyle{unsrtnat}
\bibliography{perturbation}

\pagebreak
\end{document}

%% file: defs.tex
\newcolumntype{C}[1]{>{\centering\arraybackslash}m{#1}}

\newcommand{\refalg}[1]{Algorithm \ref{#1}}
\newcommand{\refeqn}[1]{Equation \ref{#1}}
\newcommand{\reffig}[1]{Figure \ref{#1}}
\newcommand{\reftbl}[1]{Table \ref{#1}}
\newcommand{\refsec}[1]{Section \ref{#1}}

\newcommand{\etal}{$et\, \,  al.$~}

\newcommand{\add}[1]{\textcolor{red}{#1}\typeout{#1}}
\newcommand{\remove}[1]{\sout{#1}\typeout{#1}}

\newcommand{\m}[1]{\mathcal{#1}}
\newcommand{\bmm}[1]{\bm{\mathcal{#1}}}
\newcommand{\real}[1]{\mathbb{R}^{#1}}
\newcommand{\method}{\textsc{LCGCN}}
\newcommand{\methodfull}{\textsc{Landscape Crystal Graph Convolution Network}}

\newtheorem{theorem}{Theorem}[section]
\newtheorem{claim}[theorem]{Claim}

\newcommand{\reminder}[1]{\textcolor{red}{[[ #1 ]]}\typeout{#1}}
\newcommand{\reminderR}[1]{\textcolor{gray}{[[ #1 ]]}\typeout{#1}}

\newcommand{\Real}{\mathbb{R}}

\newcommand{\tuples}{\mathbb{T}}

\newcommand{\argmax}{arg\,max}

\newcommand\norm[1]{\left\lVert#1\right\rVert}

\newcommand{\note}[1]{\textcolor{blue}{#1}}

\newcommand*{\Scale}[2][4]{\scalebox{#1}{$#2$}}%
\newcommand*{\Resize}[2]{\resizebox{#1}{!}{$#2$}}%

\def\mat#1{\mbox{\bf #1}}

%% file: sections/introduction.tex
\section{Introduction}
The development of next generation technologies in energy, chemicals, electronics and transportation industries are closely tied to the development of new materials with the desired properties~\cite{sagotra2019influence}. For example in case of energy, the need of the hour is the development of inexpensive and environmentally safe materials that can generate and store electrons with very high density and efficiency while operating at high power. In case of chemicals, there is a critical need to develop new materials with better catalytic performance~\cite{eisa2022role}, as well as those which can make chemicals from alternative feedstock such as recycled plastics.

Historically, discovery of materials has been driven by empiricism, through the knowledge and intuition of researchers working on these problems. Over the past few decades, they are being augmented by the insights and knowledge derived from physics based models, in particular, atomistic models and simulations that can predict the various properties of interest for a wide range of materials~\cite{national2011materials, kumar2018machine}. However, these physics based models are computationally expensive and as a result, only aid us to explore a small subset of the materials search space. Further, within this approach, there is no formal way to transfer information and knowledge from one system to another, and from one application to another. Data driven models based on statistics and machine learning can help in speeding up the materials discovery process by providing a framework where algorithms can learn from existing data obtained from models and experiments on a subset of materials and predict the properties and functionalities of new materials. In order to facilitate such developments, various research groups from across the world have also provided free public access to the datasets of material structures ~\cite{materials_project,curtarolo2012aflow,ComputationalMaterialsDatabase,Mannodi-KanakkithodiRationalCoDesignPolymer2016,SaalMaterialsDesignDiscovery2013} and properties obtained predominantly from atomistic simulations.

These datasets and other datasets obtained from experiments have been used in the recent literature to develop data driven  models to predict material properties~\cite{pilania2013accelerating,ward2016general,kim2016organized,cgcnn}. Most of these models are based on  hand crafted descriptors that are suitable only for certain subset of materials and for targeted properties of interest. However, a transferable model needs to operate on raw data to remove these biases introduced through hand crafted descriptors. In case of materials and molecules, such raw input data are the atoms and their position coordinates. This basic materials information can be naturally expressed in the form of graphs. Graph Neural Networks (GNNs) are a class of deep learning models that have been successfully used to model graph-structured data \cite{gcn_iclr14,gdl17,Kipf2016}. GNNs can learn to effectively represent a molecular/crystal graph from raw data of atom/bond features and use neural networks to perform regression and classification tasks \cite{DuvenaudMAGHAA15,gcn_camd16,mpnn_icml17,lddd_acs17}. 
Recently, Xie \etal \cite{cgcnn} developed crystal graph convolution neural network (CGCNN) that was able to take this basic structural and chemical information of a crystal as graph to perform graph convolutions on these graphs. They were able to obtain excellent results to predict properties  of a wide range of crystals, in their ground-state equilibrium structure. Later, Sanyal \etal \cite{mtcgcnn} integrated CGCNN with multi-task learning to further improve the prediction accuracy.


%% file: sections/outline.tex
\section{Outline of the Work}

Although CGCNN has been used to predict equilibrium properties of materials, many thermodynamic, kinetic and transport properties at non-zero temperatures are determined only after incorporating the phononic contributions that distort the crystal and its atomic components from the equilibrium position. These distortion and the resultant phononic contributions depend upon the potential energy surface (PES) associated with it. PES is a highly non-linear complex function of the chemical composition, symmetry and atomic arrangement of the crystals. Even a small change in these parameters can significantly change the PES. The PES is the input to various statistical approaches such as Monte Carlo and molecular dynamics simulations  to predict the phononic degrees of freedom and their influence on various material properties.

Conventionally PES are fitted to explicit function forms from ab-initio simulation results. However, over the past few years gaussian process based machine learning has been applied to develop interatomic potentials for various materials ranging from elemental (amorphous carbon, graphene,  boron)~\cite{deringer2017,rowe2018,deringer2018} to multi-elemental (methane)~\cite{veit2019}. Although very successful, development of Gaussian Approximation Potential (GAP)~\cite{bartok2009}  requires careful design of experiments to simulate small systems whose inputs are needed to fit the interatomic potentials. Further, there are also challenges in transfering the potential to a new system.

These shortcomings can potentially be overcome by GNN based deep learning approach, in which a model learns from the raw materials data. These models can possibily lead to better transferability across different materials composition, crystal symmetry, atomic arrangements and environments. In this work, we provide an initial demonstration of such transferability and propose \methodfull{} (\method{}), an extension of CGCNN architecture, to predict the PES of alumina $\left(Al_2O_3\right)$,  a material that serves as the base material for various adsorbent and catalytic applications~\cite{hlavay2005determination,gupta2011synthesis,pines1960alumina,trueba2005gamma}. Specifically, we demonstrate the generalizability of this approach by developing \method{} models that not only accurately predict PES, but are also transferable across scales, environmental conditions (temperature) and different crystallographic polymorphs,  trigonal phase (space group R$\overline{3}c$), cubic phase (space group Ia3), Orthorhombic (space groups Pbca and Pna2$_{1}$) of alumina.



%% file: sections/method.tex
\section{Methodology}
\subsection{Molecular Dynamics}
\label{sec:md}
 The data required to train, validate and test the CGCNN models were obtained by performing Molecular Dynamics (MD) simulations. We have performed MD simulations under $( N, V, T )$ conditions  with the LAMMPS code \cite{lammps}. In these simulations the temperature is kept fluctuating around a setpoint value by using Nos\'{e} - Hoover thermostats. 
We employ  simulation boxes, typically varying from $30$ to $480$ atoms, and apply periodic boundary  conditions along the three Cartesian directions. Newton's equations of motion are integrated using the customary Verlet's algorithm with a  time-step length of $10^{-3}$~ps. The typical duration of a MD run is of $10$~ps. We used interatomic potentials of the Second-Moment tight-Binding-QE$_{q}$ (SMTB-Q) form~\cite{salles16}.

In this work, employing the above MD protocol, we simulate $Al_2O_3$ systems comprising of different sizes, environmental temperatures and polymorphs. We pick 10,000 configurations of these systems after thermal equilibriation. Unless specified otherwise, the 10,000 configurations were divided into a ratio of 6:2:2 for training, validation and testing of the \method{} model. 

\begin{figure*}
	\includegraphics[width = 0.8\linewidth]{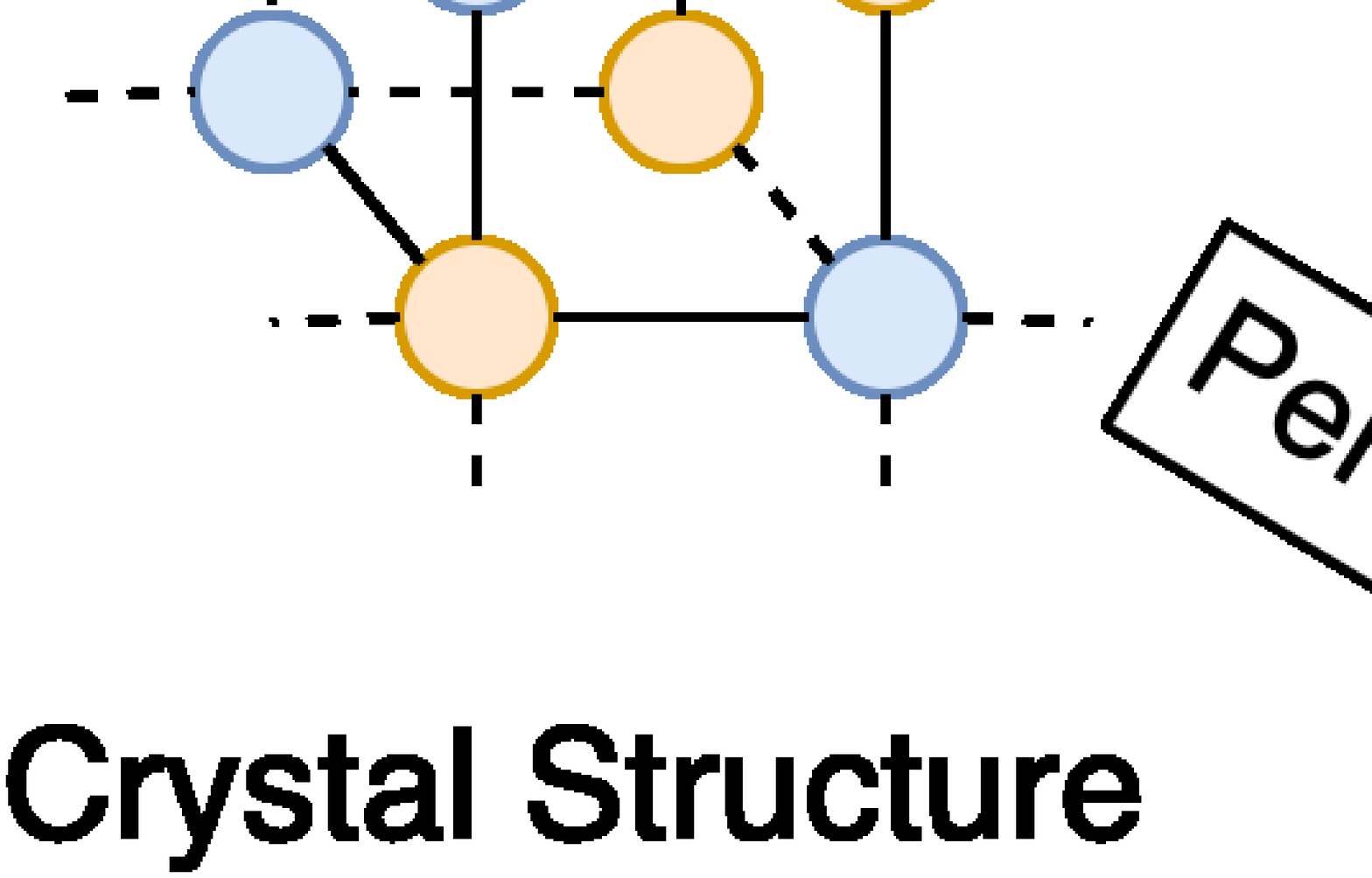}
	\caption{\label{fig:cgcnn_schematic} Schematic representation of using \method{} to predict potential energy of distorted alumina crystals. Initially, the crystal structure is represented as a graph with nodes and edges representing the atoms and bonds in the crystal respectively. Here, $\bm{v_i}, \bm{v_j}$ and $\bm{u_{ij}}$ denote the node, neighbor and edge embeddings respectively. Then, CGCNN is used to learn a crystal level embedding $\bm{v}_\mathcal{G}$ which is used to predict the potential energy. Please refer to Sec. \ref{sec:method_details} for more details. }
\end{figure*}

\subsection{\methodfull{}(\method{})}
\label{sec:method_details}
CGCNN \cite{cgcnn} is a graph neural network which focuses on building a crystal graph from a given crystal structure and utilizes graph convolution networks (GCNs) to model the crystal graph and predict their properties with accuracy comparable to \textit{ab initio} physics models. We propose \methodfull{} (\method{}), an extension of CGCNN, that can be used to predict the potential energy surface (PES) of crystal structures. Next, we discuss the details of proposed method \method{} in detail.

A crystal graph $\mathcal{G}$ is an undirected multigraph defined by nodes representing atoms and edges representing bonds in a crystal. Formally, let $\mathcal{G} \mathord{=}(\mathcal{A},\mathcal{E}, \mathcal{V}, \mathcal{U})$ denote a crystal graph. Here $\mathcal{A}$ represents the set of atoms in the crystal structure, $\mathcal{E}\mathord{=} \{(i,j)_k\mathord{:}~k^{th}~\text{bond between atom}~i~\text{and}~j \}$, is the set of undirected edges denoting the bonds and $|\mathcal{A}|\mathord{=}N$ is the number of atoms in the crystal graph. $\bm{v_i} \in \mathcal{V}$ contains the features of the $i^{th}$ atom which encodes various properties of the atom. $\bm{u}_{(i, j)_k} \in \mathcal{U}$ is the feature vector for the $k^{th}$ bond between atoms $i$ and $j$ that captures properties of the edges. To capture the influence of local neighborhood in the crystal structure, a graph convolution formulation similar to CGCNN \cite{cgcnn} is used defined as follows,

%

\small
\begin{equation*}
\label{eq:conv}
\bm{v}_i^{(t+1)} = \bm{v}_i^{(t)} + \sum_{j, k} \sigma(\bm{z}_{(i,j)_k}^{(t)} \bm{W}_c^{(t)} + \bm{b}_c^{(t)}) \odot g(\bm{z}_{(i,j)_k}^{(t)} \bm{W}_s^{(t)} + \bm{b}_s^{(t)}),
\end{equation*}
\normalsize

where $\bm{z}_{(i,j)_k}^{(t)} = \bm{v}_i^{(t)} \oplus \bm{v}_j^{(t)} \oplus \bm{u}_{(i, j)_k}$ denotes the concatenation of atom and bond feature vectors of the neighbors of $i^{th}$ atom. $\odot$ denotes element-wise multiplication and $\sigma$ denotes a sigmoid function. The $\sigma(\bm{\cdot})$ factor acts as a learned weight matrix to incorporate different interaction strengths between neighbors. $\bm{W}_c^{(t)}$, $\bm{W}_s^{(t)}$, $\bm{b}_c^{(t)}$ and $\bm{b}_s^{(t)}$ are the convolution weight matrix, self weight matrix, convolution bias and self bias of the $t$-th layer of GCN respectively, and $g(\cdot)$ is the activation function for introducing non-linear coupling between layers.

Graph pooling \cite{graphsage} is a technique used to learn graph representations given the learnt node and edge features. The learnt atom features are pooled to get a vector representation of the crystal $\left(\bm{v}_\mathcal{G}\right)$.  The crystal representation $\bm{v}_\mathcal{G}$ is then fed to a network of fully-connected layers with non-linearities that learns to predict a property of the crystal in a supervised manner. The workflow is depicted in Fig. \ref{fig:cgcnn_schematic}. To train the network, we calculate the loss using mean squared error between the true values obtained using MD simulations and the model predictions followed by back-propagation \cite{backprop}. The accuracy of the model is quantified through mean absolute error per atom (MAE/atom) by comparing the model predictions on test configurations against the true values obtained using MD simulations.


%% file: sections/results.tex
\section{Results and Discussion}

\begin{table}[]
	\centering
	\begin{tabular}{lc}
		\hline
		Pooling Operator	& Formulation		\\
		\hline
		\textsc{Average}	& $\frac{1}{N}\sum_{i=1}^N \bm{v_i}$											\\
		\textsc{Max}		& $max(\bm{v_i}), \ \ \forall i \in [1, N]$										\\
		\textsc{FC-Max}		& $max(\bm{v_i} \bm{W_{pool}} + \bm{b_{pool}}), \ \ \forall i \in [1, N]$		\\
		\hline
	\end{tabular}
	\caption{\label{tab:pool_op} Formulation of different graph pooling operators. Here, $\bm{v_i}$ denotes the atom representation and $max$ is an elementwise max operator. $\bm{W_{pool}}$ and $\bm{b_{pool}}$ are learnable pooling parameters. Please refer to Sec. \ref{sec:increase_sample} for more details.}
\end{table}

\subsection{Effect of different pooling operators}
\label{sec:pool}
In this section, we explore some of the existing graph pooling operators - \textsc{Average}, \textsc{Max} and \textsc{FC-max}. In \textsc{Average} pooling, the graph embedding is defined as the mean of all the node embeddings. Similarly, for \textsc{Max} pooling, it is defined as the elementwise max of all the node embeddings. Hamilton \etal~\cite{graphsage} proposed \textsc{FC-max}, a learnable pooling operator, which transforms node embeddings using a fully-connected neural network and then applies \textsc{Max} pooling operator. The formulation of these pooling operators are defined in Table \ref{tab:pool_op}. To analyze the effect of pooling, we perform the experiment where we use alumina trigonal polymorph data for training and predict performance on other alumina polymorphs. The results are shown in Table \ref{tab:pool}. We observe that \textsc{FC-max} performs the best among all of the pooling operators. And hence, we use \textsc{FC-max} as the pooling operator in all further analysis of our \method{} architecture.

\begin{table*}[t!]
	\centering
	\begin{tabular}{m{10em}ccc}
		\hline
		Pooling Operator	& Trigonal			& Orthorhombic (Pbca)			& Orthorhombic (Pna2$_{1}$)	\\
		\hline
		\textsc{Average}	& 0.0009			& 0.0349						& 0.1796					\\
		\textsc{Max}		& 0.0007			& 0.4256						& 0.0789					\\
		\textsc{FC-max}		& \textbf{0.0007}	& \textbf{0.0112}				& \textbf{0.047}			\\
		
		\hline
	\end{tabular}
	\caption{\label{tab:pool} MAE/atom of \method{} model for different pooling operators on test data of 240 atom alumina orthorhombic (Pbca) and Orthorhombic (Pna2$_{1}$) polymorph trained on trigonal data. Please refer to Sec. \ref{sec:pool} for details.}
\end{table*}

\begin{table*}[t!]
	\centering
	\begin{tabular}{lC{6em}C{8em}C{6em}C{8em}}
		\hline
		&\multicolumn{2}{c}{CGCNN} 			&\multicolumn{2}{c}{\method{}}		\\
		Phase						& Error/atom	& Relative Error	& Error/atom	& Relative Error	\\
		\hline
		Trigonal					& -0.0466		& 0					& -0.0018		& 0					\\
		Orthorhombic (Pbca)			& -0.0208		& 0.0258			& 0.0001		& 0.0019			\\
		Orthorhombic (Pna2$_{1}$) 	& -0.0302		& 0.0164			& 0.0031		& 0.0049			\\
		\hline
	\end{tabular}
	\caption{\label{tab:equilibrium} Error/atom (eV/atom) and Relative Error (w.r.t. trigonal alumina phase) of \method{} and CGCNN model on test dataset of different polymorphs using models trained on setups as defined in Sec. \ref{sec:equilibrium}. We observe that \method{} consistently outperforms CGCNN for all alumina phase equilibrium predictions.}
\end{table*}

\subsection{Prediction of energy at equilibrium}
\label{sec:equilibrium}
Xie \etal \cite{cgcnn} train their CGCNN model to predict formation energy per atom $\left(\Delta E_f\right)$ using a large number of equilibrium crystal structures obtained from the Materials Project \cite{materials_project}. Since $\left(\Delta E_f\right)$ for a large number of crystals can span across a wide range (approximately 8.5 eV/atom), such a model will not be able to accurately predict the small energy differences across different polymorphs of the same composition. However, in this work we train \method{} model to predict the PES of various polymorphs of alumina through training on distorted input structures obtained from MD simulations. 
The results are shown in table \ref{tab:equilibrium}. Due to this training data, we expect our model to perform better in predicting the energy differences of the alumina polymorphs. The errors of predictions from the two approaches are shown in table \ref{tab:equilibrium}. We report both 'signed' error and relative error. For relative error, we choose the energy of the trigonal phase as the reference value. As expected, \method{} significantly outperforms the previous CGCNN model {\color{black} suggesting that \method{} is well-equipped than CGCNN to predict small energy changes typically seen in distortions}. In the next upcoming subsections, we will discuss the PES predictions of alumina across supercell sizes, temperature and polymorphs.

%

\begin{figure}[h!]
	\includegraphics[width = 1.0\linewidth]{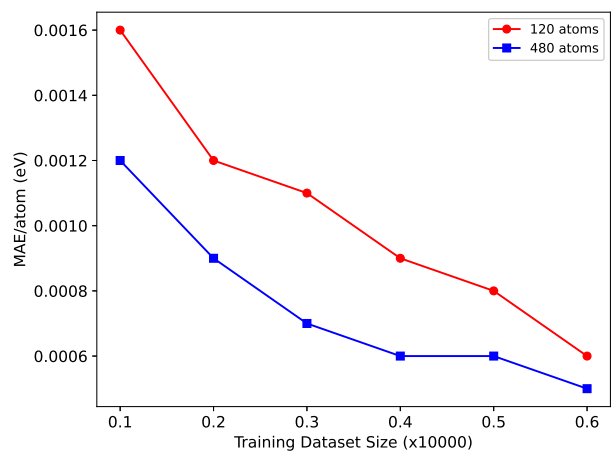}
	\caption{\label{fig:al_tri_sample} Plotting test MAE/atom of \method{} model in predicting potential energy of 120 and 480 atoms systems of alumina trigonal polymorph with increasing size of training dataset. Please refer to Sec. \ref{sec:increase_sample} for more details.}
\end{figure}
\subsection{Prediction of potential energy}
\label{sec:increase_sample}
The \method{} model was used to predict the potential energy of the trigonal alumina polymorph of different supercell sizes -- namely 120 and 480 atoms. The data to train, validate and test was obtained from MD simualtions as described in Sec. \ref{sec:md} under a temperature ramp of 300K-500K. We quantify the error by reporting mean absolute error per atom (MAE/atom) (eV). The results for the two different supercells are shown in Fig. \ref{fig:al_tri_sample}. As expected, increasing the amount of training data reduces the error. We observe that with 0.6 times the total alumina data (i.e. 6000 samples) for training, our model is able to achieve an MAE/atom which is close to the accuracy of DFT (1 meV/atom). 

\begin{table*}[t!]
	\centering
	\begin{tabular}{m{6em}m{6em}m{6em}m{6em}c}
		\hline
		\multicolumn{2}{c}{Training} 	&\multicolumn{1}{c}{Validation}			& Testing 	&		\\
		120 atoms	& 480 atoms	& 480 atoms	& 480 atoms	& MAE/atom	\\
		\hline
		1000		& 2400		& 600		& 2000		& 0.0017	\\
		3000		& 2400		& 600		& 2000		& 0.0017	\\
		6000		& 2400		& 600		& 2000		& 0.0011	\\
		10000		& 2400		& 600		& 2000		& 0.0011	\\
		\hline
	\end{tabular}
	\caption{\label{tab:al_tri_scale_exp1} MAE/atom of \method{} model on test data of 480 atoms with increasing training dataset of 120 atoms. Please refer to Sec. \ref{sec:transfer_scale} for details.}
\end{table*}

\begin{table*}[t!]
	\centering
	\begin{tabular}{m{6em}m{6em}m{6em}m{6em}m{6em}c}
		\hline
		\multicolumn{2}{c}{Training}&\multicolumn{2}{c}{Validation}	& Testing 		&			\\
		120 atoms		& 480 atoms	& 120 atoms			& 480 atoms	& 480 atoms		& MAE/atom	\\
		\hline
		2400			& 0			& 600				& 0			& 2000			& 0.0023	\\
		2400			& 800		& 600				& 200		& 2000			& 0.0014	\\
		2400			& 1600		& 600				& 400		& 2000			& 0.0013	\\
		2400			& 2400		& 600				& 600		& 2000			& 0.0011	\\
		\hline
	\end{tabular}
	\caption{\label{tab:al_tri_scale_exp2} MAE/atom of \method{} model on test data of 480 atoms with increasing training data of 480 atom configurations. Please refer to Sec. \ref{sec:transfer_scale} for further details.}
\end{table*}

\subsection{Transferable \method{} models across scale}
\label{sec:transfer_scale}
\begin{figure}[t!]
	\includegraphics[width = 1.0\linewidth]{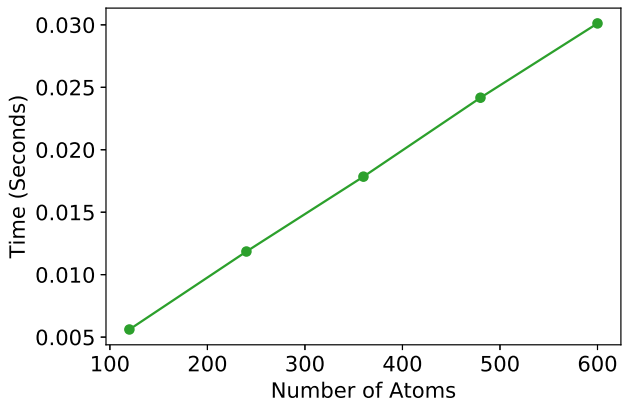}
	\caption{\label{fig:time_scale} Time analysis of the \method{} code with respect to the CPU run, as a function of number of atoms.}
\end{figure}

In many scenarios, the property of interest can only be obtained from performing large scale atomistic and quantum mechanical simulations. However, the scaling of such simulations with system size can be non-linear. For example in case of DFT, the cpu time of a simulation increases as $O(N^3)$ where $N$ is the number of atoms that limits the system size. Hence, an ML model that is transferable from smaller scale systems to larger scale systems are of importance in the field. Firstly, we note that the cpu time for inference using our model increases as $O(N)$ with the number  of nodes $N$. We perform a time analysis of the \method{} model. For this, we use a pre-trained \method{} model to predict potential energy for a set of 100 configurations of 120 atom alumina trigonal polymorph. Next, we repeat this for 240, 360, 480 and 600 atoms supercells. The average cpu time for each setup is depicted in  Fig. \ref{fig:time_scale}. We observe that \method{} is significantly faster than DFT and MD simulations, and in contrast to MD and DFT the LCGCN scales linearly with the system size and easily parallelized \citep{smith2017ani,  behler2008pressure}.

\begin{figure}[t!]
	\includegraphics[width = 1.0\linewidth]{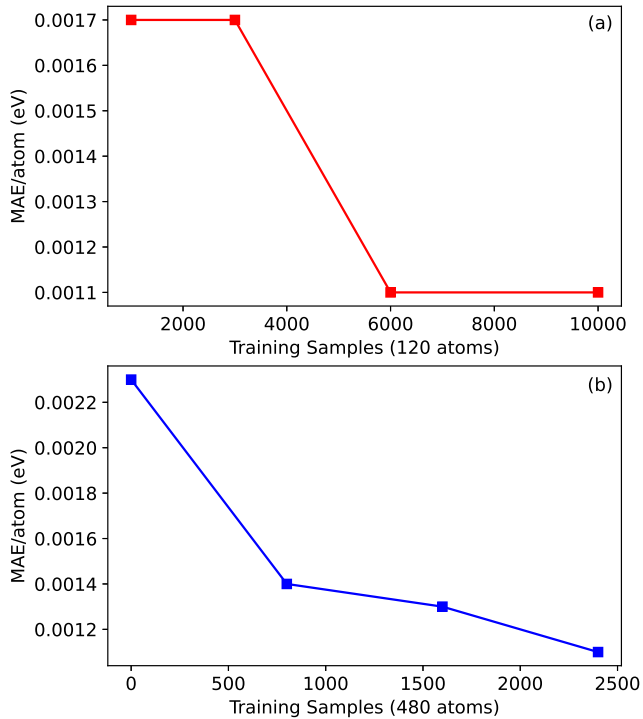}
	\caption{\label{fig:al_tri_scale1} (a) Test MAE/atom of \method{} model in predicting potential energy of 480 atoms of alumina trigonal polymorph as a function of increasing training dataset of 120 atoms. Refer Sec. \ref{sec:transfer_scale} for details. (b) Test MAE/atom of \method{} model in predicting potential energy of 480 atoms of alumina trigonal polymorph as a function of increasing number of 480 atom configurations in training dataset. Refer to Sec. \ref{sec:transfer_scale} for details.}
\end{figure}

Next, we demonstrate how increasing the training data of the 120 atoms supercell of alumina trigonal polymorph can reduce the test error of predicting the potential energy of the 480 atom supercell configurations. For this, we consider a setup where there are a fixed 2400 configurations of 480 atoms for model training. Also, we have 10000 configurations of 120 atoms that we progressively add to the training set. As observed from Table \ref{tab:al_tri_scale_exp1}, an increase in the number of 120 atom configurations in training data leads to a reduction in the test error for the 480 atom configuration. After adding all the 10000 configurations of 120 atoms in training data, the error in MAE/atom for 480 atom configuration is 0.0011 eV/atom. This is shown in Fig. \ref{fig:al_tri_scale1}a. In a second set of experiments, the data of 120 atoms is fixed in training set, while the number of 480 atoms is increased in training. The results are reported in Table \ref{tab:al_tri_scale_exp2} and depicted in Fig. \ref{fig:al_tri_scale1}b. We find that an increase in the 480 atom configuration in training set leads to significant improvements in model performance on predicting potential energy for 480 atoms. From these set of experiments we can conclude that although it is not trivial for \method{}  to generalize, although with comparitively higher error to predict PES of the larger supercell (480 atoms) even when these larger supercell configurations are not included in the training data (1st row of Table \ref{tab:al_tri_scale_exp2}). However, this error goes down significantly even if a small amount of the 480 atoms configurations are included into the training data (4th row of Table \ref{tab:al_tri_scale_exp2}).


\subsection{Transferable \method{} models across environmental conditions}
\label{sec:transfer_env}
Envrionmental conditions, in particular temperature, play an important role in determining the dynamics of the materials and the resultant phase stability and material properties~ {\color{black} \cite{sagotra2019influence, klarbring2018phase}}. Higher temperature molecular dynamics has been employed as a tool in different problems to sample the configurations rapidly and to predict the resultant low energy configuration and their corresponding properties {\citep{behler2007generalized}. In this section, we analyze \method{}'s transferability in predicting potential energies of configurations of trigonal alumina across different environmental conditions. We specifically focus on variation of temperature and test the effectiveness of the model in three temperature values for alumina - 300K, 500K and 800K. We pretrain a \method{} model using all the data (10,000 configurations) from one temperature and predict potential energy at a different temperature. The results are shown in Table \ref{tab:al_temp_exp1}.

{\color{black} In $\alpha$-Al$_{2}$O$_{3}$, two types of Al-O bonds (1.87 angstrom and 1.98 angstrom) are present at 0 K. But under classical MD simulation, these two bonds vary drastically at a given temperature T. At low temperature (say 300 K), bond-bond distribution function shows two peaks (at 1.85 and 2.03 angstrom) while at high temperatures (500 K $\&$ 800 K), the peak corresponding to longer (weaker) Al-O bond (2.03 angstrom peak in 300 K configuration) gets disappeared, see Fig. \ref{fig:bond_dist}a. It means that weakly bonded Al and O got separated much compared to strongly bonded Al and O due to thermal stress.}

Next, we observe the effect of mixing the data across two different temperatures (resulting in a total of 20,000 configurations split into training, validation and test set) in predicting energies for data at an unseen temperatures. The results are shown in Table \ref{tab:al_temp_exp2}. We see when the model is fed with a mixture of two temperatures data for training the prediction error reduces significantly for test data at unseen temperatures. Specifically, we observe a significant decrease in the MAE for prediction at 800K, when data for both 300K and 500K are mixed. In order to delineate if the improvement in model prediction is due to training data volume, or due to data from two different temperatures, we train a model with 5000 configurations of 300K and 500K each and test the model on data from 800K. The resulting MAE/atom ($4^{th}$ row of Table \ref{tab:al_temp_exp2}) is still significantly lower than predictions at 800K using only 300K or 500K data for training. This suggests that the model is able to learn better when it is trained with heterogeneous data from two different temperatures.

\begin{table}[t!]
	\centering
	\begin{tabular}{m{4em}m{5em}m{5em}m{5em}}
		\hline
		Temp.		& 300K		& 500K		& 800K		\\
		\hline
		300K		& 0.0003	& 0.0212	& 0.0766	\\
		500K		& 0.0151	& 0.0004	& 0.0254	\\
		800K		& 0.0372	& 0.0192	& 0.0007	\\
		\hline
	\end{tabular}
	\caption{\label{tab:al_temp_exp1} MAE/atom of \method{} model on test data of 240 atom alumina trigonal polymorph at three different temperatures. Row temperature indicates that model was trained using the data of that particular temperature and the column temperature denotes that the trained model was used to predict potential energy at that temperature. Please refer to Sec. \ref{sec:transfer_env} for more details.}
\end{table}

\begin{table}[t!]
	\centering
	\begin{tabular}{m{12em}m{5em}m{5em}}
		\hline
		Mixture					& Target	& MAE/atom		\\
		\hline
		300K \& 500K			& 800K		& 0.02			\\
		300K \& 800K			& 500K		& 0.0048		\\
		500K \& 800K			& 300K		& 0.0035		\\
		300K \& 500K(5000 each)	& 800K		& 0.0169		\\
		
		\hline
	\end{tabular}
	\caption{\label{tab:al_temp_exp2} MAE/atom of \method{} model on test data of 240 atom alumina trigonal polymorph at a target temperature using a mixture of data for two different temperatures in training data. Please refer to Sec. \ref{sec:transfer_env} for details.}
\end{table}

\begin{figure}[t!]
	\includegraphics[width = 1.0\linewidth]{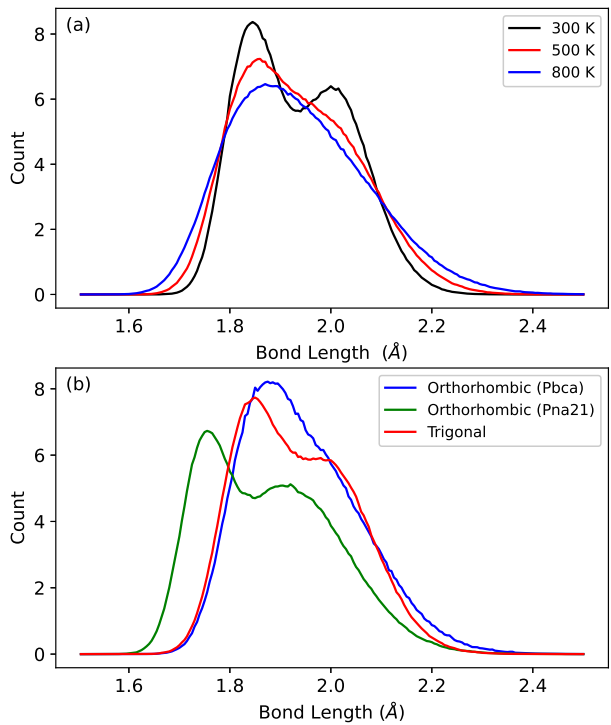}
	\caption{\label{fig:bond_dist} (a) Al-O bond distribution of $\alpha$-Al$_{2}$O$_{3}$ at 300 K , 500 K and 800 K temperatures.(b)Al-O bond distribution of Alumina polymorphs. (c) Density of energy distribution of Alumina polymorphs.}
\end{figure}

\begin{table*}[t!]
	\centering
	\begin{tabular}{lccc}
		\hline
		Phase 						& Orthorhombic (Pbca)  	& Trigonal 	& Orthorhombic (Pna2$_{1}$) \\
		\hline
		Orthorhombic (Pbca)    		& 0.0007 				& 0.0265  	& 0.037 					\\
		Trigonal 					& 0.0112 				& 0.0007   	& 0.047 					\\
		Orthorhombic (Pna2$_{1}$)   & 0.9689 				& 0.1542   	& 0.0006 					\\
		\hline
	\end{tabular}
	\caption{\label{tab:al_poly_exp1} MAE/atom of \method{} model on test data of 240 atom alumina for three different polymorphs. The row phase denotes that the model was trained using the data for that particular polymorph and column phase indicates that the trained model was used to predict the potential energy for that polymorph. Refer to Sec. \ref{sec:transfer_poly} for more details.}
\end{table*}

\begin{table}[]
  	\centering
  	\begin{tabular}{lc}
	   	\hline
	    Mixture & MAE/atom \\
	    \hline
		Trigonal + Orthorhombic (Pna2$_{1}$)(5000 each) 	& 0.1804          \\
	    Trigonal + Orthorhombic (Pna2$_{1}$)    			& 0.1312          \\
	    \hline
	\end{tabular}
  	\caption{\label{tab:al_poly_exp2} MAE/atom of \method{} model on test data of 240 atom alumina othorhombic (Pbca) polymorph using a mixture of data for different alumina polymorphs in training data. Please refer to Sec. \ref{sec:transfer_poly} for details.}
\end{table}

\subsection{Transferable \method{} models across different Polymorphs}
\label{sec:transfer_poly}
One of the major challenges in materials and chemistry arises from the difficulty in transfering information, model and knowledge across different crystal structures. A model that can learn properties such as potential energy, from equilibrium and distorted configurations of some crystal structures and predict for new crystal structures with minimal data would be a very useful tool in accelerating materials discovery. As a preliminary step toward this goal, in this work, we study the transferability of \method{} models across different polymorphs of alumina. We consider three phases of alumina - orthorhombic (Pbca and Pna2$_{1}$) and trigonal. For the first set of experiments, we predict the potential energy of the different polymorphs based on the training data from another polymorph (\textit{aka} zero shot learning). Similar to temperature study, here we use all the data pertaining to a polymorph (10,000 configurations) to train the model. The results are shown in Table \ref{tab:al_poly_exp1}. 

To understand the trends in the transferability of \method{} across different polymorphs, we plotted the $Al-O$ bond distribution and configurational energies for the three alumina polymorphs as depicted in Fig. \ref{fig:bond_dist}b respectively. We observe that, the more the overlap between $Al-O$ bond distribution of different polymorphs, the better transferability, and hence lower the error in \method{}'s prediction. For example, the overlap between the Orthorhombic (Pbca) and Trigonal polymorphs is greater than that between the Orthorhombic (Pbca) and Orthorhombic (Pna2$_{1}$) polymorphs. As seen in Table \ref{tab:al_poly_exp1}, our \method{} model, when trained on Orthorhombic (Pbca) data, predicts the Trigonal polymorph with a lower error compared to when it is trained on Orthorhombic (Pna2$_{1}$). This is likely due to the fact that the \method{} model's accuracy is highly dependent on bond length, and transferability is better when predicting systems with similar bond lengths.




In order to understand how the model accuracy improves when trained with heterogeneous data, we train the model with data from multiple polymorphs to predict potential energy of orthorhombic (Pbca) polymorph. We choose orthorhombic (Pbca) for our analysis because the error in PES prediction was the highest as shown in Table \ref{tab:al_poly_exp1}. The results for this analysis is shown in Table \ref{tab:al_poly_exp2}. We observe that if we restrict the training data to a total of 10,000 configurations (each polymorph has 5,000 data points), the error in prediction is about 0.1804 eV/atom suggesting that training on diverse dataset can improve the transferability of the model. However, if we incorporate all the data from the two polymorphs (20,000 configurations in total), the model error further reduces to 0.1312 eV/atom suggesting that the transferability of the model can be improved by increasing both the volume and variety of training data.

%% file: sections/conclusion.tex
\section{Conclusion}
{\color{black} In this work, we proposed \method{}, a graph neural network framework to directly predict  high-dimensional potential energy surface for alumina $\left(Al_2O_3\right)$, a base material for various adsorbent and catalytic applications, from atomic configurations. This approach enables to bypass calculating the derivatives of the PES which are often the computational bottlenecks of existing {\textit{ab-initio}} methods.

We further demonstrated the transferability of \method{} across scales, environmental conditions and different alumina polymorphs. Finally, We validated that, with appropriate conditioning, \method{} can obtain accuracies that are close to DFT (1 meV/atom). Therefore, we anticipate that \method{} will significantly accelerate the  atomistic studies of other materials  with its scalability, versatility, and excellent potential energy surface prediction accuracy.}

\section*{DATA AVAILABILITY}
The data that support the findings of this study are available from the corresponding author (J.B. and A.K.S) upon
reasonable request.

%% file: sections/acknowledgement.tex
\section*{ACKNOWLEDGMENTS}
The authors would like to acknowledge Shell, JNCASR and IISc for their support. \\ 

\section*{AUTHOR CONTRIBUTIONS}
J.B, U.V.W and P.T conceived the study and planned the research. A.K.S. and N.K. performed the MD simulations to generate the dataset. S.S , M. K,  S. R and A.K.S implemented and optimized the \method{} model. All authors discussed the results and their implications and contributed to the writing of the article. \\


\section*{COMPETING INTERESTS}
The authors declare no competing interests.